\documentclass[prb,twocolumn,showpacs]{revtex4}
\usepackage{times}
\usepackage{amsmath}
\usepackage{amsthm}
\usepackage{amssymb}
\usepackage{bm}
\usepackage{natbib}
\usepackage{amsbsy}
\usepackage{graphicx}
\usepackage{pstricks}
\usepackage{color}
\begin{document}

\pdfoutput -1

\newcommand{\hexa}{\;
\pspicture(0,0.1)(0.35,0.6)
\psset{unit=0.75cm}
\pspolygon(0,0.15)(0,0.45)(0.2598,0.6)(0.5196,0.45)(0.5196,0.15)(0.2598,0)
\psset{linewidth=0.08,linestyle=solid}
\psdots[linecolor=gray,dotsize=.20](0,0.15)
\psdots[linecolor=gray,dotsize=.20](0.2598,0.6)
\psdots[linecolor=gray,dotsize=.20](0.5196,0.15)
\endpspicture\;}

\newcommand{\hexb}{\;
\pspicture(0,0.1)(0.35,0.6)
\psset{unit=0.75cm}
\pspolygon(0,0.15)(0,0.45)(0.2598,0.6)(0.5196,0.45)(0.5196,0.15)(0.2598,0)
\psset{linewidth=0.08,linestyle=solid}
\psdots[linecolor=gray,dotsize=.20](0,0.45)
\psdots[linecolor=gray,dotsize=.20](0.5196,0.45)
\psdots[linecolor=gray,dotsize=.20](.2598,0)
\endpspicture\;}

\newcommand{\hexdb}{\;
\pspicture(0,0.1)(0.35,0.6)
\psset{unit=0.75cm}
\psset {linewidth=0.03,linestyle=solid}
\pspolygon(0,0.15)(0,0.45)(0.2598,0.6)(0.5196,0.45)(0.5196,0.15)(0.2598,0)
\psset{linewidth=0.1,linestyle=solid}
\psline(0.2598,0.6)(0,0.45)
\psline(0.5196,0.15)(0.5196,0.45)
\psline(0.2598,0)(0,0.15)
\endpspicture\;}

\newcommand{\hexda}{\;
\pspicture(0,0.1)(0.35,0.6)
\psset{unit=0.75cm}
\pspolygon(0,0.15)(0,0.45)(0.2598,0.6)(0.5196,0.45)(0.5196,0.15)(0.2598,0)
\psset{linewidth=0.1,linestyle=solid}
\psline(0,0.15)(0,0.45)
\psline(0.2598,0.6)(0.5196,0.45)
\psline(0.5196,0.15)(0.2598,0)
\endpspicture\;}

\newcommand{\smallhexh}{ \;
\pspicture(0,0.1)(0.2,0.3)
\psset{linewidth=0.03,linestyle=solid}
\pspolygon[](0,0.0775)(0,0.225)(0.124,0.3)(0.255,0.225)(0.255,0.0775)(0.124,0)
\endpspicture
\;}

\begin{abstract}
We study a model of strongly correlated spinless fermions on a kagome lattice at 1/3 filling, with interactions described by an extended Hubbard Hamiltonian. An effective Hamiltonian in the desired strong correlation regime is derived, from which the spectral functions are calculated by means of exact diagonalization techniques. We present our numerical results with a view to discussion of possible signatures of confinement/deconfinement of fractional charges.

\end{abstract}

\title{Strongly correlated fermions on a kagome lattice}
\author{A. O'Brien}
\address{Max Planck Institute for the Physics of Complex Systems, 38 N{\"o}thnitzer Str.,01187 Dresden, Germany}
\address{Institut f{\"u}r Physik, Technische Universit{\"a}t Chemnitz, D-09107 Chemnitz, Germany}
\author{F. Pollmann}
\address{Department of Physics, University of California, Berkeley, CA94720, USA}	
\author{P. Fulde}
\address{Max Planck Institute for the Physics of Complex Systems, 38 N{\"o}thnitzer Str.,01187 Dresden, Germany}
\address{The Asia Pacific Center for Theoretical Physics, Pohang Korea}
\maketitle

\section{Introduction}\label{intro}
Excitations with fractional charges have for a number of years been a subject of intense investigation in condensed matter physics. This strong interest was initially triggered by the work of Su, Schrieffer and Heeger \cite{Su80},
 who showed that in trans-polyacetylene such excitations may exist when the system is doped. \cite{Su81} The fractional charges depend on doping concentration and are especially pronounced for certain $\pi$ band fillings. Their appearance does not require electron-electron interactions but lattice degrees of freedom play an important role via symmetry breaking. Even before the work of Su, Schrieffer and Heeger it was realised by Jackiw and Rebbi \cite{Jackiw76}, that a system with Dirac particles and a symmetry breaking term gives rise to excitations with charge $e/2$. In the mean time the same type of theory was applied to graphene \cite{Chamon08} and to a kagome lattice with $1/3$ $(2/3)$ filling. \cite{Franz09}

A quite different origin of excitations with fractional charges was discovered by Laughlin \cite{Laughlin83} with his explanation of the fractional quantum Hall effect. Here electron-electron interactions are crucial. In fact, it is required that the system is in the limit of strong electronic correlation where the kinetic energy of the electrons plays a minor role in comparison to the Coulomb repulsion. The elimination of the kinetic energy is caused by a strong external magnetic field which forces the electrons into the lowest Landau level. In distinction to the former case, lattice degrees of freedom are not important here. It was also found that fractional charges obey fractional statistics. \cite{Wilczek84}

Subsequently it was shown that there exists a further class of systems with strong electron correlations which yield excitations with fractional charges $\pm e/2$. Prerequisites are a geometrically frustrated lattice structure like a pyrochlore, checkerboard \cite{Runge04} or kagome lattice as well as certain lattice fillings. The excitations with fractional charges can be either confined \cite{Pollmann06b} or deconfined.\cite{Sikora09} They are low-energy in nature and can be treated in the absence of magnetic excitations (for instance through the employment of doped dimer models as in this paper \cite{poilblanc08}). It is interesting that a case of deconfined excitations with fractional charges on a three dimensional (3D) pyrochlore lattice can be identified. This calls into question relationships between fractional charges and fractional statistics since in 3D there are only fermions or bosons possible but not anyons. In contrast, the 2D case readily admits the possibility of fractional charges with anyonic statistics. Direct comparison of model systems for both bosons and fermions \cite{Ralko} may shed insight on the nature of the statistics of emergent fractional charges.\\

Emerging experimental techniques lead us to consider also how such problems might be investigated in ultracold atomic gases; for example, a recent proposal in the form of tunable optical lattice schemes would be readily applicable to our model. \cite{Ruostekoski} In the context of synthesised materials, spinel compounds contain kagome planes and are therefore natural candidates for possible experimental verification (a material of particular pertinence here is discussed in Ref.~\onlinecite{Kamimura}); there is furthermore already experimental evidence that electrons in pyrochlore lattices can be strongly correlated. \cite{Kondo,Walz}

In the present communication we wish to elaborate on the kagome lattice at $1/3$ and $2/3$ fillings. Since we are interested only in charge degrees of freedom, rather than spin degrees of freedom, we consider spinless fermions, or their equivalent, fully spin polarized electrons. In that case a lattice site is empty or singly occupied but never doubly occupied due to the Pauli exclusion principle. 
We assume again that the electron correlations are strong. We also show that, upon the addition or removal of a particle, vacuum fluctuations occur. Each of these particles therefore decays into two `defects', each with a fractional charge of $\pm$$\frac{e}{2}$. The theory breaks the symmetry between the positively and negatively charged excitations. The paper focuses on the spectral density and its properties. In particular we want to show how the strong correlations and fractional charges show up in that quantity. The hole spectral function has symmetry properties which are unraveled by gauge transformations. It also has features which originate from eigenstates of an effective Hamiltonian. The latter is obtained in the strongly correlated limit whereby nearest neighbour hopping processes are replaced by ring hopping which lifts the otherwise macroscopic degeneracy of the ground state.

The paper is organized as follows. In Section II, we introduce the model Hamiltonian. Both the non-interacting and interacting cases are considered. An effective Hamiltonian valid in the limit of strong electronic correlations is derived and found to map directly to a quantum dimer model with resonating plaquettes. \cite{Moessner01c}Symmetries of this effective model are explained and discussed. In Section III, the single-particle spectral functions are defined and numerical calculations using the exact diagonalization method are presented.

Finally Section IV provides an overview and conclusion. Appendix A provides a proof of a generalized sign rule which applies in the limit of strong correlations. 

\section{Model Hamiltonian}\label{sectwo}

We start from an extended Hubbard model on a kagome lattice
with nearest-neighbor repulsion $V$. As we are working with spinless fermions, i.e., fully spin polarized electrons, 
no on-site repulsion term is necessary. Using second quantized notation, the Hamiltonian is written as
\begin{equation}
H=-t\sum_{\langle
i,j\rangle}\left(c_{i}^{\dag}c^{\vphantom{\dag}}_{j}
+ \text{H.c.}\right) \\
  +V\sum_{\langle
i,j\rangle}n_{i}n_{j}.
\label{eq:extended_hub}
\end{equation}
Here, the operators $c^{\vphantom{\dag}}_{i}$ ($c_{i}^{\dag}$)
annihilate (create) a fermion on site $i$. The number operators are given by $n_{i}=c_{i}^{\dag}c^{\vphantom{\dag}}_{i}$. The notation
$\langle i,j\rangle$ refers to pairs of nearest neighbors. 

\subsection{Non-interacting case}
\begin{figure}[t]
  \centering
  \includegraphics[width=3in]{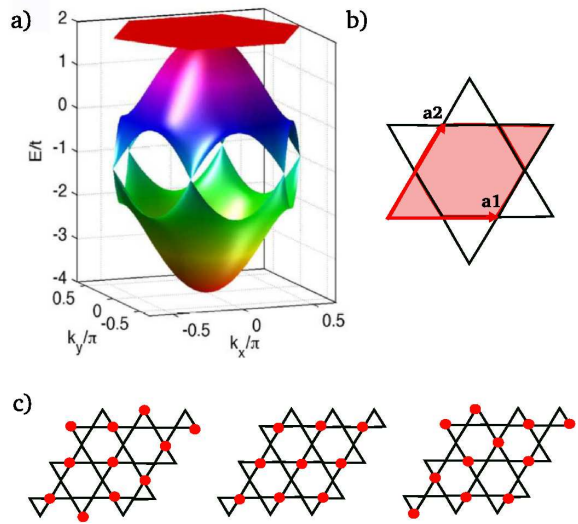}
  \caption[Non doped kagome lattice at 1/3 filling]
  {(color online) a) The three bands of the kagome lattice are shown in the first Brillouin zone.  b) The lattice vectors $\mathbf{a}_1$ and $\mathbf{a}_2$ are shown on a section of the kagome lattice; they trace out the unit cell, of the lattice. c) Three possible configurations that minimise the potential energy at 1/3 filling (these constitute just a small fraction of possible ground state configurations for this cluster size).} 
    \label{fig:intro}
\end{figure}

If no interactions are present, i.e., $V=0$ in Hamiltonian Eqn.~(\ref{eq:extended_hub}), we can simply diagonalize the Hamiltonian in momentum space. The spectrum consists of three bands. \cite{Franz09} One of them is a flat band $\epsilon_k^1=2t$ and the other two are dispersive: 
\begin{equation}
\epsilon_k^{2,3}=t\left[-1 \pm \sqrt{4(\cos^{2}k_1+\cos^{2}k_2+\cos^{2}k_3)-3}\right],\label{eq:disp}
\end{equation}
where $k_n=\mathbf{k}\cdot\mathbf{a}_n$ are given with respect to the lattice vectors $\mathbf{a}_1=\hat{x}$, $\mathbf{a}_2=(\hat{x}+\sqrt{3}\hat{y})/2$, $\mathbf{a}_3=\mathbf{a}_2-\mathbf{a}_1$. The lattice vectors are defined as depicted in FIG.~\ref{fig:intro}b. 

FIG.~\ref{fig:intro}a  shows the three bands in the first Brillouin zone. The three-fold band structure has a lowest energy level with a minimum energy of $-4t$. The band picture demonstrates the nature of the ground state at 1/3 filling: only the first band is filled, so that the Fermi level sits exactly at the Dirac points. This property can lead to interesting physics if the Hamiltonian is weakly perturbed as shown in Ref.~\onlinecite{Franz09}.

\begin{figure}[t]
  \centering
  \includegraphics[width=3in]{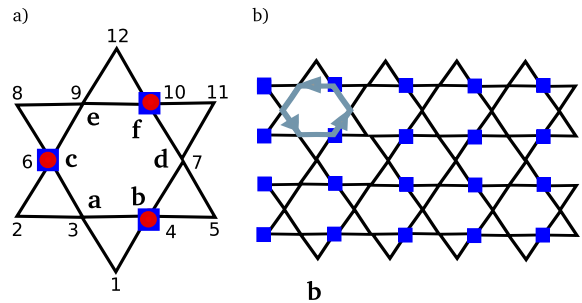}
  \caption[row enumeration of the kagome lattice]
  {(color online) a) One possible ring exchange process is shown here, where particles on sites $f$, $b$ and $c$ can hop anticlockwise (clockwise) to sites $e$ ($d$), $d$ ($a$) and $a$ ($e$) respectively. Enumerating row-wise as shown, such a process involves always an odd number of anticommutations between particles.
  b) Particle hopping around a hexagon as shown always results in a change of the number of particles on the sublattice of two; the local gauge transformation derivable from this picture explains the symmetry of $H_{\text{eff}}$.
}
\label{fig:rowenum}
\end{figure} 

\subsection{Strong interactions}
In this paper we consider mainly the model at 1/3 filling in the limit of strong correlation ($t/V \ll 1$). For strongly correlated fermion models on certain frustrated lattices it has been shown that at low energies fractionalization of charge does occur. \cite{Fulde02} In order to study certain effects of that fractionalization, we derive an effective Hamiltonian which acts on the low-energy subspace of the full Hilbert space in the strong correlation regime. 

A similar technique has been applied to the extended Hubbard model on the checkerboard lattice. \cite{Runge04,Pollmann06c,trousselet2008}

We first derive an effective Hamiltonian for the exactly 1/3 filled case (``undoped case'') and then consider the case in which one particle is added or removed from the system (``doped case''). Using the effective model, we are able to perform exact diagonalizations of clusters up to 108 sites.   
\paragraph*{Undoped case:}  In order to derive an effective model for the quantum strong correlation limit, we start from the limit in which $t/V=0$. A value of $V > 0$ will give rise to a local constraint, the triangle rule. Configurations  $|C\rangle$ which fulfill this constraint have exactly one fermion on each triangle; in this case the mutual repulsion energy is zero. FIG.~\ref{fig:intro}c shows different degenerate ground-state configurations. The number of degenerate ground states of the system scales exponentially as a function of system size. 

As we turn on a small $|t|\ll V$, the macroscopic degeneracy is lifted. The lowest-order term in $t/V$ that lifts the degeneracy stems from ring-hopping processes around hexagons (indeed, ring exchange processes are found in various frustrated spin systems \cite{Diep,Pollmann06b}). All lower order terms lead solely to a constant energy shift and do not affect the low-energy physics.
The low-energy excitations, up to third order in $t/V$, are described by the following effective Hamiltonian
\begin{equation}
H_{\text{eff}}= g \sum_{\smallhexh}\left( | \hexa \rangle  \langle \hexb |+\mbox{H.c.}\label{eq:heff}\right).
\end{equation}
This Hamiltonian acts within the manifold of configurations which fulfill the triangle rule  and the effective ring-hopping amplitude is given by $g=12t^3/V^2$. The growth with system size here is also exponential, albeit much slower than that in the case of the extended Hubbard model. An approximate formulation of this growth is given in Ref.~\onlinecite{Pauling} as $(\frac{4}{3})^{\frac{N}{2}}$ where N is the number of occupied sites (particles). An exact value for the bulk entropy per particle in the thermodynamic limit given in can be compared with the corresponding finite cluster quantities.\cite{Elser} 

Even though the original Hamiltonian in Eqn.~(\ref{eq:extended_hub}) is fermionic, the effective Hamiltonian in Eqn.~(\ref{eq:heff}) has no sign change in the matrix elements.  Indeed, for all hexagonal flipping processes it can be shown that

\begin{equation}
\langle C'| \hexa \rangle \langle  \hexb | C \rangle \rightarrow -1. 
\label{eq:hexflip}
\end{equation}

This remarkable property belonging to all matrix elements of the effective Hamiltonian can be obtained through a row-wise enumeration of the kagome lattice sites. With such enumeration, we find that an odd number of anticommutations must be performed by the $c$-operators in (3) when evaluating the matrix elements, thus always contributing a negative sign. This can be seen by considering a ring exchange process on a hexagon as labeled in FIG.~\ref{fig:rowenum}a. In this case, one particle will always move to an adjacent site (thereby undergoing no anticommutations which could introduce a sign). The remaining two particles in the process both anticommute with a nonzero number of particles. Crucially, one must undergo exactly one more (or less) anticommutation then the other. Hence, the two particle processes \emph{always} involves an odd number of anti-commutations and thus the sign of any such ring exchange process is negative, as in Eqn.~(\ref{eq:hexflip}). The argument holds for any hexagon on a row-wise enumerated kagome lattice. From this observation it follows that the spinless fermion effective model at exactly $1/3$ filling is equivalent to a hard-core bosonic model. Furthermore, the effective model at$1/3$ filling is equivalent to that at $2/3$ filling.

We remark that the overall sign in Eqn.~(\ref{eq:heff}) is just a matter of a simple
local gauge transformation.
Specifically, the sign of $g$ in Eqn.~(\ref{eq:heff}) can be changed by
multiplying all configurations with a phase 
\begin{equation}
|C\rangle \rightarrow i^{\nu(C)}|C\rangle,
\label{eq:hexgauge}
\end{equation}
where $\nu(C)$ is the number of particles on the sublattice shown in
FIG.~\ref{fig:rowenum}b. This fact might appear surprising since the actual sign of $g$ can typically be gauged away only for the
cases of bipartite lattices.
The sign of $g$ turns out
to be inconsequential due to the constrained nature of the ring exchange quantum dynamics of
Eqn.~(\ref{eq:heff}) and is a desirable feature. A system so described by matrix elements which are all non-positive can be studied through Monte Carlo simulation. It is also interesting for physical reasons as it implies that the system must have a symmetric eigenspectrum. 

  \begin{figure}[ht]
  \centering
  \includegraphics[width=3in]{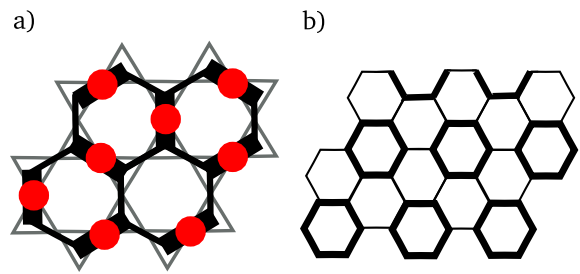}
  \caption[Non doped kagome lattice at 1/3 filling]
  {(color online) a) The honeycomb lattice (black) comprises of the links of the medial lattice belonging to the kagome lattice (light grey). Occupied sites of the kagome lattice map to dimers on the honeycomb lattice. b) Schematic diagram of the dimer representation of the so-called ``plaquette phase'' in which dimers resonate around the bold hexagons.}

  \label{fig:intro2}
\end{figure}
Configurations $|C\rangle$ on the kagome lattice fulfilling the triangle rule correspond to dimer configurations on the honeycomb
lattice (particles are placed here on links, see FIG.~\ref{fig:intro2}a. Different dimer coverings are orthogonal because any wavefunction overlap is neglected. In the dimer representation, the effective Hamiltonian reads
\begin{equation}
H_{\text{QDM}}=g\sum_{\smallhexh}\left(|\hexda \rangle  \langle \hexdb |+H.c.\right).
\end{equation}
The phase diagram of the QDM on a honeycomb lattice is to a large extent already understood. \cite{Moessner01c} The case which we consider here, i.e., that of $H_{\text{QDM}}$, has a ground state which corresponds to the three-fold degenerate plaquette phase with broken translational invariance. \cite{Moessner01c, Isakov06a}  A schematic of this phase is shown in FIG.~\ref{fig:intro2}b. Importantly, the separation of a pair of defects introduced into this phase tends to destroy the plaquette order. This, in turn, reduces the number of ring exchange processes and hence increases the energy of the system. Hence, $g$ acts on pairs of defects as a confinement potential; this intriguing feature of the plaquette phase results in a ground-state for the systems that is both charge ordered and confined. The dynamical nature of the confinement potential is investigated in Section \ref{spectra} of this paper.

\paragraph*{Doped case:}
\begin{figure}[t]
\centering
\includegraphics[width=3in]{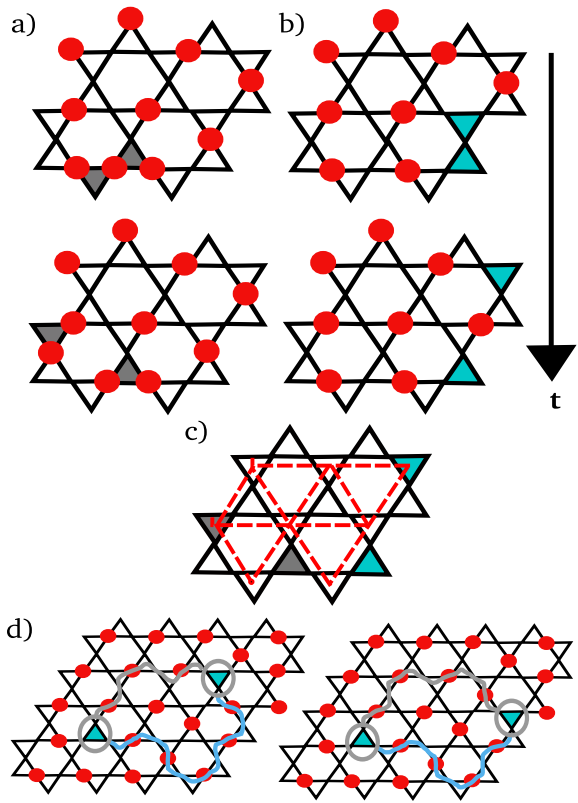}
\caption{(color online) a) Charge fractionalization due to \emph{adding} one particle to the 1/3 filled lattice. b) Charge fractionalization due to \emph{removing} one particle from the 1/3 filled lattice. c) Each particle or hole excitation creates two defects. Each of these lives on one of the two sublattices formed by triangles which constitute the kagome lattice (that is, on a triangular sublattice, like the example shown). d) Here one sees two configurations which are connected by a matrix element of the hopping operator. The parity of the particle number $d_{\mathcal{L}}(C)$ on any connecting path $\mathcal{L}$ with alternating empty and occupied sites between the two defects is always even in one case and always odd in the other case (two possible paths are shown in each case).}
\label{fig:charges}
 \end{figure}
 
After the effective low-energy description for the undoped case, we now turn to the case where the system is doped, i.e., a particle is either removed (hole-doping) or added (particle doping) to a system at $1/3$ filling. 

Placing one additional particle with charge $-e$ onto an empty site
of a configuration which fulfills the triangle rule leads to a violation on two adjacent triangles, see FIG.~\ref{fig:charges}a.
The energy is increased by $2V$ since the added particle has two
nearest neighbors (charge gap). There is no way to remove the violations
of the triangle rule by moving particles. However, particles on triangles with two particles can hop to another neighboring
triangle without creating additional violations
of the triangle rule, i.e., without increase of the repulsion energy
(see FIG.~\ref{fig:charges}a). By these hopping
processes, two local defects (=violations of the triangle rule)
can separate and the added particle with charge $-e$ breaks into two
pieces. They carry a fractional charge of $-e/2$ each. 
In the quantum mechanical case ($t\not=0$), the separation leads
to a gain in the kinetic energy of order $t$. A similar scenario takes places when removing one particle from a configuration which fulfills the triangle rule. Removing a particles creates two adjacent defects with no particle on a triangle. These defects can separate and each of them is carrying a charge of $+e/2$, see FIG.~\ref{fig:charges}b. 

The dynamics of the particle or hole doped system is described by the effective Hamiltonian Eqn.~(\ref{eq:heff}) with an added hopping term:
\begin{eqnarray}
H_{\text{doped}}=&g& \sum_{\smallhexh}\left( | \hexa \rangle  \langle \hexb |+\mbox{H.c.}\right)-\nonumber\\
&t&\sum_{\langle i,j\rangle}P(c_i^{\dag}c_j^{\vphantom{\dag}}+H.c.)P.
\label{eq:doped}
\end{eqnarray}
Here, the operator \emph{P} serves to project out high-energy states, i.e., it projects out all configurations with additional violations of the triangle rule (as we consider only configurations with one pair of defects here). 

For the effective Hamiltonian Eqn.~(\ref{eq:heff}), we showed above that the fermionic sign does not influence the low energy excitations and that the sign of the effective ring-exchange amplitude $g$ is inconsequential. However, as soon as the system is doped a fermionic sign problem arises (as the second term in (\ref{eq:doped}) does not necessarily preserve the ring-exchange symmetry of the first term). It turns out that the projected hopping term of defects in the hole doped system (i.e., Hamiltonian Eqn.~(\ref{eq:doped}) with $g=0$) is also bipartite. This remarkable feature of the hole-doped hopping counters what we would naively expect. The defects hop on two triangular sublattices of the honeycomb lattice (see FIG.~\ref{fig:charges}c) which are \emph{not} bipartite. However, the bipartite nature of this hopping term is in fact a consequence of the triangle rule to which it is subject; the nature of the restricted hopping is clearly distinct from the free hopping of particles on the triangular sublattices. Furthermore, the sign of $t$ can be changed by simply multiplying all configurations $|C\rangle$ with a phase
\begin{equation}
 |C\rangle \rightarrow (-1)^{d_{\mathcal{L}}(C)}|C\rangle,\label{eq:sign_doped}
 \end{equation}
 where $d_{\mathcal{L}}(C)$ is the number of particles on an arbitrary path $\mathcal{L}$ between the two defects (see FIG. ~\ref{fig:charges}d), which has alternating occupied and empty sites. Note that for finite systems with periodic boundary conditions an extra phase also emerges if a particle hops around the torus. A detailed derivation of the sign rule is given in Appendix~\ref{sign_rule}. \\
 
The projected hopping term as well as the ring exchange term for hole defects possesses such a symmetry. However, the combination of the two, Hamiltonian $H_{\text{doped}}$ (\ref{eq:doped}), does not generally possess a symmetric eigenspectrum (as the ring exchange term symmetry only exists in the absence of the hopping term). That is, the symmetry no longer holds for the doped system. The reason is that the dynamics of a doped hole involves even and odd permutations of fermions, which destroy the parity conservation of the sublattice shown in FIG.~\ref{fig:rowenum}b. Consequently in the presence of hole hopping processes the ring exchange term of the effective Hamiltonian is no longer bipartite. 
 \begin{figure*}[t]
  \centering
  \includegraphics[width=7.5in]{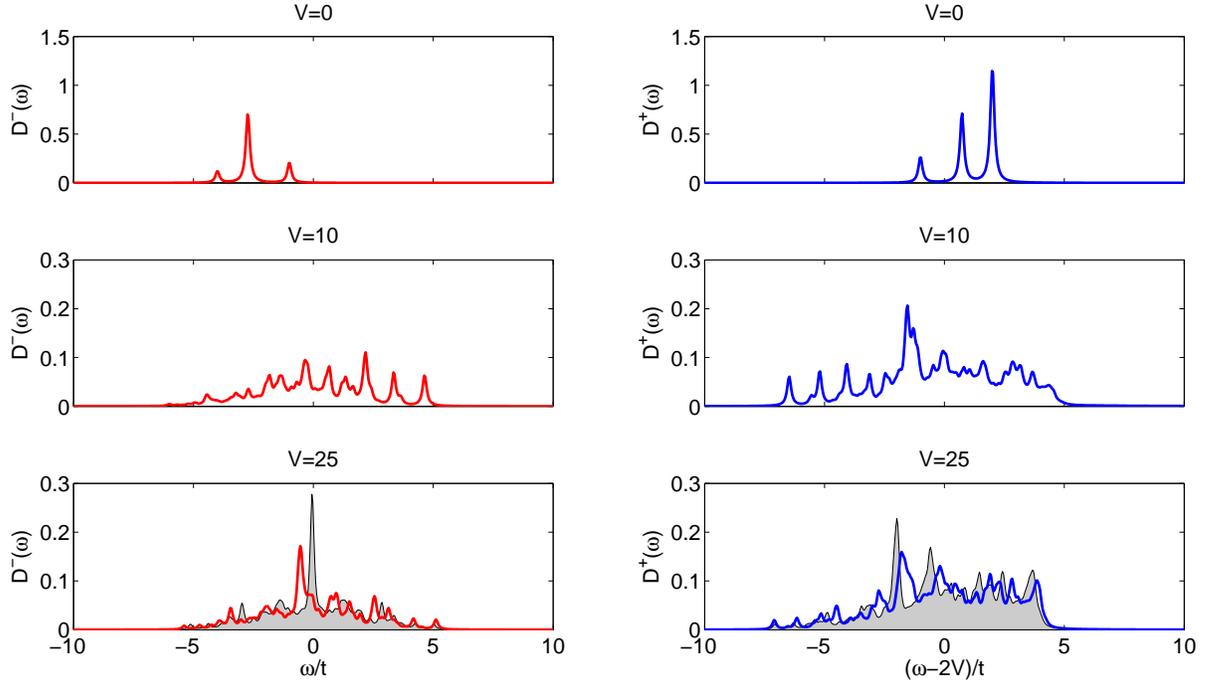}	
  \caption{(color online) The DOS for several ratios of $V$ and $t$ for the extended Hubbard model is shown above. A small cluster with $27$ sites allows for this full treatment. For ease of comparison, the effective DOS is also plotted for $V=25$, i.e., in the strongly correlated regime (shaded in grey, it can be seen to closely resemble the full DOS as expected).}
  \label{fig:tvFigure}
\end{figure*}

\section{Spectral Functions}\label{spectra}

The dynamics of the fractional charges can be studied through calculation of the spectral function $A(\mathbf{k},\omega)=A^{+}(\mathbf{k},\omega)+A^{-}(\mathbf{k},\omega)$ \cite{Pollmann06e}, for which the particle and hole contributions are:

\begin{eqnarray}
A^{+}(\mathbf{k},\omega)&=&\lim_{\eta \rightarrow 0^{+}} -\frac{1}{\pi} \times \\
& &\sum_{a=1}^3\text{Im}\langle \psi_0^N|c^a_{\mathbf{k}}\frac{1}{\omega+i\eta+E_0-H}(c^a_{\mathbf{k}})^{\dag}|\psi_0^N \rangle\nonumber \\
A^{-}(\mathbf{k},\omega)&=&\lim_{\eta \rightarrow 0^{+}} -\frac{1}{\pi} \times \\
 & &\sum_{a=1}^3\text{Im}\langle \psi_0^N|(c^a_{\mathbf{k}})^{\dag}\frac{1}{\omega+i\eta-E_0+H}c^a_{\mathbf{k}}|\psi_0^N \nonumber\rangle
\end{eqnarray}
Here, $|\psi_0^N\rangle$ is the ground state of a N particle system with a corresponding energy $E_0$. The operator $c_{\mathbf{k}}^{a\vphantom{\dag}}$ annihilates a particle with momentum $k$ in band $a$ of the Hamiltonian in the non-interacting case and is obtained from the corresponding operators in real-space representation by Fourier transformation. The spectral functions are directly measured in ARPES experiments and are hence a promising candidate for yielding experimental signatures of fractionalization.
 
 Recent studies showed that the quantity $A^{+}(\mathbf{k},\omega)$  may also be accessible through photoemission spectroscopy undertaken in optical lattice experiments. \cite{Jin}
  
  The resulting integrated spectral density is \begin{eqnarray}
D^{\pm}(\omega)=\frac{1}{N}\sum_{\mathbf{k}}A^{\pm}(\mathbf{k},\omega).\end{eqnarray}
For a system with a translationally invariant ground state,  the quantity $D(\omega)$
can be conveniently calculated in real-space representation from local creation and annihilation operators. \\

We first consider the DOS for the extended Hubbard model; this full treatment is possible as we use our smallest cluster with $27$ sites. Tuning through the various regimes of the model for this quantity, it is possible to follow the transition from non-interacting case to the strongly correlated case. We can furthermore demonstrate the near equivalence of the effective model in the regime of strong correlations ($V=25t$ tunes the model to well within this regime).
In FIG.~\ref{fig:tvFigure} the DOS is thus shown for different values of nearest-neighbor repulsion $V$. In the non-interacting case, $V=0t$, the spectral functions can be calculated directly from the dispersion relation Eqn.~(\ref{eq:disp}). The contributions result from the allowed k-points in the considered cluster. The bandwidth is $6t$, in agreement with that seen in FIG.~\ref{fig:intro}.\\

Moving away from the non-interacting limit, we find that a more complex structure emerges. A gap opens between the hole and particle contributions; this is due to the $2V$ energy contribution the addition of a particle adds to the system. The gap opening signifies a metal-insulator transition and is estimated to appear at $V \approx 3t$. \cite{Nishimoto}
The delta peaks give way to broadened features, characteristic of the high density of low-lying excitations that arise in the presence of repulsive interactions. This is shown for $V=10t$. This feature as well as a broadened bandwidth, become more marked with increasing repulsion energy strength, until we approach the limiting case of strong correlations ($V$ is effectively infinite with respect to $t$  within this regime and hence all values of $V/t$ here yield the qualitative features expected when $t/V \rightarrow 0$). Here the broadened structure in the limit of strong correlations has an enhanced bandwidth much larger than $6t$, indeed it is $\approx$$12t$ if one considers just the bandwidth of the two contributions separately. This can be interpreted as a consequence of the fractionalization of the hole or particle defects that occurs in the presence of strong repulsive interactions.  Furthermore, in this limit,
the calculated spectral densities for the effective model give a valid description of the extended Hubbard model, i.e., below the threshold of $t/V<1/25$ the effective model closely approximates the extended Hubbard model. For $V= 25t$ it is seen that the effective and exact (extended Hubbard) models closely resemble each other; each peak in the exact model has a corresponding feature in the effective one. However, with respect to each other they are shown to be shifted in energy. This is primarily due to the self-energy contribution (consisting of all processes where a fermion hops to a neighbouring site, or around a triangle, then back to its original site) which is absent in the effective model calculation. This discrepancy between the two spectral functions is of order $\sim$${t^2}/{V}$ and disappears entirely as $t/V \rightarrow 0$.

\subsection{Asymptotically free case ($g \ll t$)}
\begin{figure}[t]
  \centering
  \includegraphics[width=3in]{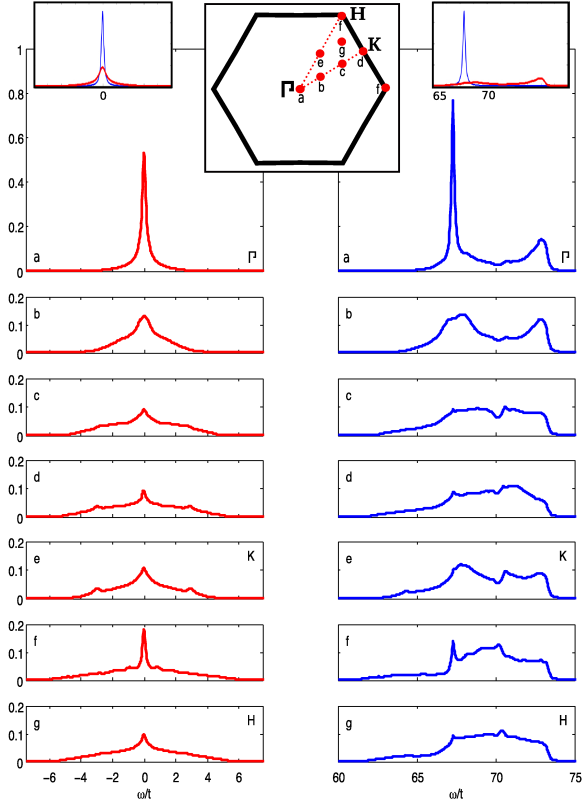}
  \caption[]
  {(color online) The Brillouin Zone is shown along with all distinct k-points for a 108-site cluster. The particle and hole contributions are shown in red (light grey) and blue (dark grey) to the left and right respectively along three lines of symmetry. The k-points in k-space have the coordinates $a=(0,0), b=(0,0.19\pi) ,c=(0,0,38\pi) ,d=(0,0.58\pi) ,e=(0.33\pi,0) ,f=(0.17\pi,0.48\pi) ,g=(0.67\pi,0)$, in units of 1/a (where lattice constant a = $|\mathbf{a}_1/2|$ = $|\mathbf{a}_2/2|$ , see FIG.~\ref{fig:intro}. Inset: the band-resolved spectral functions are shown for the gamma point; the delta peaks in both the hole and particle doped cases correspond to the lowest energy band (band 3). The lower curves represent the contributions of the other two bands 1 and 2.}
  \label{fig:figureN2}
\end{figure}

We now start by considering the effective Hamiltonian of Eqn.~(\ref{eq:doped}) in the limit of infinitesimally small $g \ll t$ (i.e., case the limit of $t/V\rightarrow0$). 
 \begin{figure}[t]
  \centering
  \includegraphics[width=8cm]{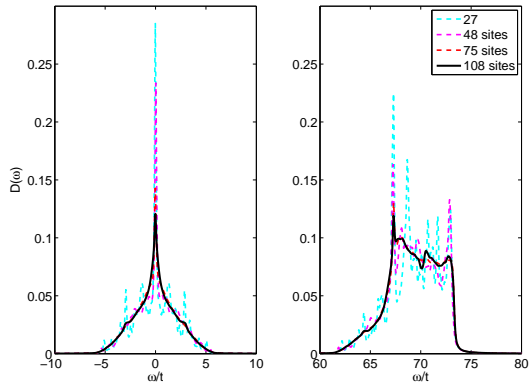}	
  \caption{(color online) The hole (left panel) and particle (right panel) contribution to the local density of states $D(\omega)$ for $g=0.01$ for various cluster sizes.} 
    \label{fig:finitesizescaling}
\end{figure}
In the effective model, $g$ mediates the plaquette ordering strength and therefore acts as a parameter for the strength of the confinement of the fractional charges. Thus infininitesimally small $g$ can be thought of as the case where the fractional charges are effectively deconfined, i.e., the spatial extent of the bound pair is effectively infinite compared to the cluster size. FIG.~\ref{fig:finitesizescaling} shows the k-integrated spectral density in this limit for various cluster sizes. As cluster size increases, key features of the DOS like bandwidth and spectral distribution are maintained. 
This is expected, as the scale of a bound pair is much larger than that of any of our clusters. 
We turn now to the angle-resolved spectra. 
In the following we present results and an analysis pertaining to these calculations, for the largest accessible cluster with $108$ sites.
The spectrum of the hole spectra is almost perfectly symmetric as shown in FIG.~\ref{fig:figureN2}. This is a numerical manifestation of the local gauge symmetry, presented in Section \ref{sectwo}, for the projected hopping of hole defects. As discussed in Section \ref{sectwo}, $H_{\text{doped}}$ does not generally possess such symmetry, but as the spectra in FIG.~\ref{fig:figureN2} pertains to the limit of very small $g$, the spectral symmetry is nevertheless preserved to very good approximation. Although not presented here it is also possible within our calculations to observe the symmetry of the ring exchange term in the undoped case numerically (recall from Eqn.~(\ref{eq:hexgauge}) that this gauge symmetry arises from the bipartiteness of the ring exchange processes).\\ 

\begin{figure}[t]
  \centering
  \includegraphics[width=3in]{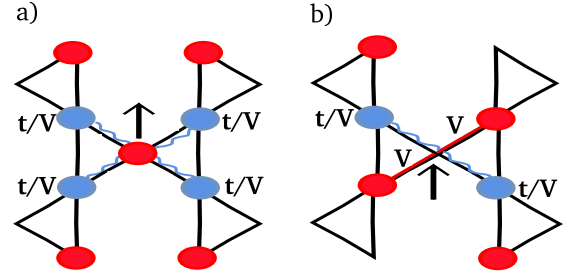}
  \caption[Quantum fluctuations at 1/3 filling]
  {(color online) a) Particle removed at 1/3 filling; due to strong repulsions $V$, such a particle is surrounded by unoccupied sites. However, due to quantum fluctuations these sites are occupied with probability $t/V$. b) Particle added at 1/3 filling; due to strong $V$ repulsions any such particle has exactly two nearest neighbours, with the probability $t/V$ of having extra nearest neighbours due to quantum fluctuations.}
  \label{fig:perturbation}
\end{figure}

The most prominent feature at the $\Gamma$-point  in both the hole and particle doped spectral functions is a sharp peak (see FIG. ~\ref{fig:figureN2}, inset). The peaks originate from the fact shown below that $|\tilde{\psi}^{N-1}\rangle=c^{3}_{\mathbf{k}=\mathbf{0}}|\psi^N_0\rangle$ and $|\tilde{\psi}^{N+1}\rangle=(c^{3}_{\mathbf{k}=\mathbf{0}})^{\dag}|\psi^N_0\rangle$ are approximate eigenfunctions of the Hamiltonian Eqn.~(\ref{eq:extended_hub}) in the limit of small $|t|/V$. 
The operator $c^{3\dag}_{\mathbf{k}=\mathbf{0}}$ creates a fermion the third band with zero momentum, in the non-interacting case; i.e., in the lowest dispersive band, as given by the dispersion relations in Eqn.~(\ref{eq:disp}). 
 The k-space and real space operators for band 3 are related by:

\begin{eqnarray}
c^{3}_{\mathbf{k=0}}=\frac{1}{\sqrt{3N}}\sum_j c_j.
\end{eqnarray}
where $c^{3}_{\mathbf{k=0}}$ annihilates a particle at site j. We now demonstrate in detail that $|\tilde{\psi}^{N-1}\rangle$ is an approximate eigenfunction in the hole doped case:
\begin{eqnarray}
H|\tilde{\psi}^{N-1}\rangle & = & Hc^{3}_{\mathbf{k}=\mathbf{0}}|\psi_{0}\rangle\nonumber \\
 & = & \left[H,c^{3}_{\mathbf{k}=\mathbf{0}}\right]|\psi_{0}\rangle+c^{3}_{\mathbf{k}=\mathbf{0}}H|\psi_{0}\rangle\nonumber \\
 & = & \left[H,c^{3}_{\mathbf{k}=\mathbf{0}}\right]|\psi_{0}\rangle+E_{0}|\tilde{\psi}^{N-1}\rangle,\end{eqnarray}
where $E_0$ denotes the ground-state energy of the undoped system. 
For the kinetic energy contribution to the
commutator we obtain the following expression:
\begin{eqnarray}
\left[H_{\mbox{\tiny kin}},c^{3}_{\mathbf{k}=\mathbf{0}}\right] & = & \sum_{a=1}^3\sum_{\mathbf{k^{\prime}}}\left[\epsilon^a_{\mathbf{k^{\prime}}}n^{a}_{\mathbf{k^{\prime}}},c^{3}_{\mathbf{k}=\mathbf{0}}\right]\nonumber \\
 & = & \epsilon^3_{\mathbf{k}=\mathbf{0}}c^{3}_{\mathbf{k}=\mathbf{0}} =  -4t\  c^{3}_{\mathbf{k}=\mathbf{0}},
 \end{eqnarray}
where we have used the dispersion relation Eqn.~(\ref{eq:disp}). The commutator involving the repulsion energy is calculated using the real space representation:

\begin{eqnarray}
\left[H_{\mbox{\tiny rep}},c^{3}_{\mathbf{k}=\mathbf{0}}\right] & = & \frac{V}{\sqrt{3N}}\sum_{\langle ij\rangle}\left(c_{j}n_{i}+c_{i}n_{j}\right).
\end{eqnarray}
In the considered limit, each site has four neighboring sites $\langle i,j\rangle$ which are occupied with probability $t/V$ due to quantum fluctuations in $|\psi_0^N\rangle$ (see FIG.~\ref{fig:perturbation}). Thus the sum over all nearest neighbors of an arbitrary site applied to the ground state leads to
\begin{eqnarray}
\left[H_{\mbox{\tiny rep.}},c^{3}_{\mathbf{k}=\mathbf{0}}\right]|\psi_{0}\rangle & = & V\frac{1}{\sqrt{3N}}\sum_{i}4\frac{t}{V}\ c_{i}|\psi_{0}\rangle\nonumber \\
 & = & 4t|\tilde{\psi}^{N-1}\rangle.\end{eqnarray}
Collecting all terms yields\begin{eqnarray}
H|\tilde{\psi}^{N-1}\rangle=\left(-4t+4t+E_{0}\right)|\tilde{\psi}^{N-1}\rangle=E_{0}|\tilde{\psi}^{N-1}\rangle.\end{eqnarray}
 Thus the peak in the spectral function $A^-(\mathbf{k}=\mathbf{0},\omega)$ is located near $\omega=0$ (note that only corrections of order $t/V$ are taken into account).

An analogous derivation for the particle doped case shows that $|\tilde{\psi}^{N+1}\rangle$ is an eigenfunction of $H$ with
\begin{eqnarray}
H|\tilde{\psi}^{N+1}\rangle&=&\left(\epsilon^3_{\mathbf{k}}+2V+2t+E_{0}\right)|\tilde{\psi}^{N+1}\rangle
\nonumber\\ 
&=&\left(2V-2t+E_{0}\right)|\tilde{\psi}^{N+1} \nonumber\rangle
\end{eqnarray}
and thus the peak in the spectral function $A^+(\mathbf{k}=\mathbf{0},\omega)$ is located near $\omega=2V-2t$. These peaks are signatures of approximate eigenstates of the full Hamiltonian in the limit of small $t$ and large $V$ and are therefore expected to arise in spectral density results for the full Hamiltonian as well as those for the effective Hamiltonian. Calculation of this quantity was also performed for the extended Hubbard model for a cluster with $27$ sites.. 
Calculations for larger clusters are presently not possible because the unrestricted Hilbert space becomes unmanageable.

As previously mentioned, a discrepancy between the effective and full models, in the position of the hole and particle peaks of order $~t^2/V$, is expected.

 \begin{figure*}[t]
  \centering
  \includegraphics[width=7in]{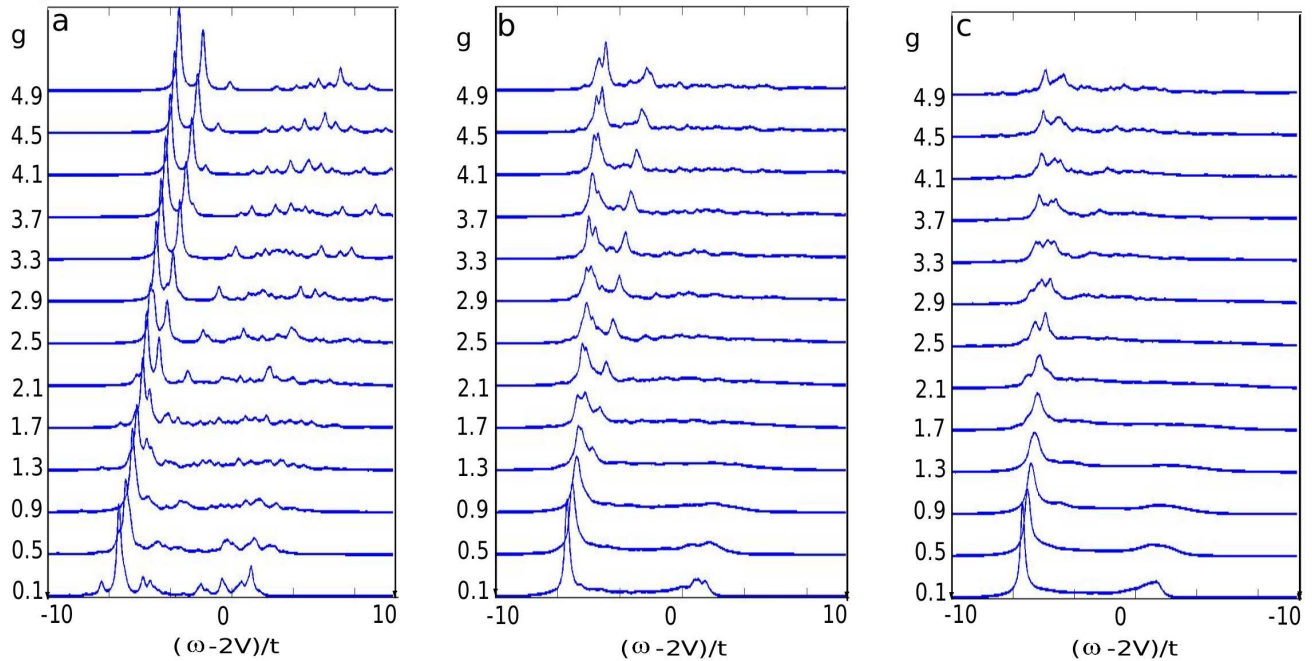}
   \caption{(color online) Here, the spectral function $A^{+}(\mathbf{k=0},\omega)$, is shown over a range of $t$ for clusters with 27 and 48  and 75 sites (shown from left to right respectively). It is estimated that at a value of $g \approx t$ the bound pairs are critically confined just within the cluster boundary. We interpret the formation of two modes at $g \approx 2t$ as a signal of this critical behaviour (see text for details).}	
  \label{fig:gtfig}
\end{figure*}

\subsection{Strongly confined case}

Consider now the behavior of the effective model where $g$ is significant compared to $t$. Crucially, the ``plaquette ordering" ring exchange term is now tuned to a value such that the fractional charges are confined at scales comparable to the cluster size.

In the thermodynamic limit, the confined nature of the defects would always be observable in the spectral densities. However, the energetic confinement of the fractional charges on the lattice is masked on a sufficiently small cluster with periodic boundary conditions (this is the case of effectively deconfined fractional charges as discussed in the previous subsection). For sufficiently large $g$ this is no longer the case; we observe bound states at the point at which the confinement strength brings the two defects just within the boundaries of the cluster. We present numerical results and analyses precisely outlining this behaviour.\\

As two defects separate from each other through hopping processes, they generate a ``string" connecting them to one another. There is an increase of kinetic energy along the string locally, as the separating defects reduce the number of flippable hexagons through their movement through the plaquette-ordered background. The resulting confinement strength between the two particles is quantified in terms of the string tension, which is the energy gradient obtained as the defects separate. The potential for a single defect is given by $U(r)=sgr$, where $s$ is the string tension, $g$ is the ring-hopping strength and $r$ is the distance between the two defects.\\

An estimate for the discrete energy levels for a bound state of two defects, is given by the solution of the 1D Schr\"{o}dinger Equation with $U(r)$ as the potential energy term. The solutions, subject to appropriate boundary conditions, are given by:
\begin{equation}
E_{N}=-\text{Root}_{\text{Airy}_N}\times(s g)^{2/3},
\label{eq:airysoln}
\end{equation}
where $\text{Root}_{\text{Airy}_N}$ is the $N$th root of the Airy function $Ai(r)$. Comparing these energy levels with values of $U(r)$ used in our numerical simulations, we estimate that for $g \approx t$, confined defects in the lowest bound state will be just inside the boundaries of our finite clusters. The string tension has been estimated through a comparison of ground-state energies for clusters with static defects separated at various distances. As expected, estimating the critical value of $g$ using (\ref{eq:airysoln}) agrees in a good approximation with results in FIG.~\ref{fig:gtfig}. Here at around $g \approx 2t$ for all three cluster sizes, two distinct modes appear for the particle contribution to the spectral function at the $\Gamma$-point. We interpret these modes as quantized bound states that appear as the confinement strength becomes sufficiently large to confine the two particles with the cluster boundaries.\\
The appearance of bound states in the spectra at a critical value of the potential is in principle observable experimentally. Indeed, analogously, bound states have been recently observed in Ising Chains in experiment. \cite{kiefer}

Lastly we comment on the bandwidths seen for the DOS.
From the spectral densities shown for smallest $g$ in FIG. ~\ref{fig:finitesizescaling} it can be seen that in both cases the bandwidths are $\approx$$12t$. 
 
This holds in the aforementioned limit of infinitesimal $g$; however, it can be seen that the bandwidths increase with increasing $g$, in accordance with the corresponding relaxation of the restricted hopping constraints (this can be observed in FIG.~\ref{fig:gtfig}, most prominently for the smallest cluster with 27 sites). 

\section{Summary and Outlook}\label{sum}
We have derived an effective model for spinless fermions at 1/3 filling in the limit of strong correlations. This has been mapped to a quantum dimer model with resonating plaquettes; subsequently a term which dopes the system with defects has been added. A local gauge symmetry has been found for $+$$e$ hole hopping. The spectral functions for both hole and particle doped systems have been calculated and this local gauge symmetry has been observed. The primary feature of the DOS has been identified as an approximate eigenstate of the full Hamiltonian in the limit of strong correlations. The confinement of the fractional charges has been confirmed qualitatively through our spectral density calculations for large values of $g$ in the effective model. In future work we intend to extend our calculations of spectral densities to include an analogous model system of hard-core bosons at 1/3 filling, with a view to understanding the nature of the highly non-trivial statistics of the fractional charges. As a realization of our model system using ultracold atoms has been proposed, a natural extension of the work would be to identify spectral features that would be observable in experiment.\\

\section{Acknowledgements}\label{sum}

We are grateful to Ari M. Turner, Andreas M. Laeuchli and Michael Schreiber for useful discussions and in particular to Nic Shannon for many discussions and valuable contributions to our understanding of spectral densities and dimer models. The work was in part supported by a grant from the U.S. National Science Foundation I2CAM International Materials Institute Award, Grant DMR-0645461.
\appendix
\section{Generalized sign rule}\label{sign_rule}
In this appendix, we prove the generalized sign rule for hole doping in the case of an infinite lattice. We first show that every loop with alternating occupied and empty sites on the kagome lattice has a length of $L=2k$ sites with $k$ being an \emph{odd} integer if the ``triangle rule'' is fulfilled. This fact is then used to directly derive the sign rule Eqn.~(\ref{eq:sign_doped}).
 \begin{figure}[t]
  \centering
  \includegraphics[width=3in]{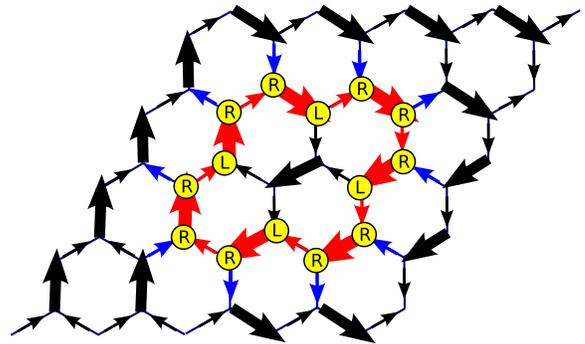}	
  \caption{(color online) Weighted arrows are used to represent a dimer configuration on the honeycomb lattice; dimers are represented by bold arrows pointing (with weight 2) and empty links by small arrows (weight 1).  Arrows of weight 2 (1) point towards A (B), one of the two bipartite sublattices. A loop with an odd number of alternating occupied and empty links is shown, along with labels for right and left turning vertices.}
  \label{fig:dimers}
\end{figure}

For convenience, we chose to represent the allowed configurations on the kagome lattice, i.e., those which fulfill the ``triangle rule'' in the dimer representation on a honeycomb lattice (see FIG.~\ref{fig:intro2}b). Since the honeycomb lattice is bipartite, we can add orientations (arrows) to the dimers. We replace each dimer with an arrow of weight 2 pointing from sublattice A to B and each empty link with an arrow of weight -1 pointing from B to A (see FIG.~\ref{fig:dimers}). In this representation, the ``triangle rule'' implies that the sum of all arrows at each vertex is zero, in other words, the ``flux'' through each vertex is conserved.  Every closed loop with alternating occupied and empty kagome sites corresponds to a loop of arrows pointing in the same direction with alternating weights  2 and -1, respectively.  As illustrated in FIG.~\ref{fig:dimers}, a loop involves $n_l$ left turns of $-60^{\circ}$ and $n_r$ right turns of $60^{\circ}$ which have to add up to $360^{\circ}$. Only at  ``right turns''  an arrow is pointing either into or out of the surface surrounded by the loop. Thus flux conservation requires that $n_r$ is even.  Since we consider loops with alternating weights on the links, the total number of links ($=n_r+n_l$) has to be even. The three conditions for $n_r$ and $n_l$ can be expressed as  
\begin{equation}
n_r-n_l=6 ,  n_r=2h , n_r+n_l=2k
\end{equation}
with $h$ and $k$ being integer. The above equations imply that $k=2h-3$ and thus the length of the loop is $L=2k$ with $k$ being \emph{odd}. This concludes the proof.  

If we now remove a particle from an allowed configuration, the two defects can only hop on loops ${\mathcal{L}}$ with alternating occupied and empty links. Since the loops contain an odd number of dimers in the undoped case (as shown above), the number of particles is now even. Thus, the number of particles which are in between the two defects is the same in either direction and  is changed by one after each hopping process. The sign of the hopping term is changed by simply multiplying all configurations $|C\rangle$ with a phase
\begin{equation}
 |C\rangle \rightarrow (-1)^{d_{\mathcal{L}}(C)}|C\rangle,
  \end{equation}
 where $d_{\mathcal{L}}(C)$ is the number of particles in between the two defects. If a finite lattice with periodic boundary conditions is considered, the transformation needs to take into account a phase factor  when loops wind around the torus.

\end{document}